\documentclass[twocolumn]{aastex701}

\usepackage{graphicx}
\usepackage{physics}
\usepackage{mathrsfs}
\usepackage{subcaption}

\newif\ifinternal
\internalfalse     

\definecolor{deeppink}{rgb}{1.0, 0.08, 0.58}

\newif\ifrev
\revfalse          

\newcommand{\rev}[1]{\ifrev\textcolor{blue}{#1}\else#1\fi}

\newcommand\bb[1]{\boldsymbol{#1}}
\newcommand\bcdot{\,\bb{\cdot}\,}
\newcommand\btimes{\,\bb{\times}\,}
\newcommand\bdbldot{\,\bb{:}\,}
\newcommand\rmd{\mathrm{d}}

\shorttitle{Reconnection-driven electron energization in KHI}
\submitjournal{ApJL}

\begin{document}

\title{Reconnection-induced electron energization in magnetospheric Kelvin--Helmholtz dynamics}

\author[orcid=0009-0001-2448-9854,sname='Silvia Ferro']{Silvia Ferro}
\affiliation{Centre for mathematical Plasma Astrophysics, Department of Mathematics, KU Leuven, Celestijnenlaan 200B, B-3001 Leuven, Belgium}
\email[show]{silvia.ferro@kuleuven.be}
\correspondingauthor{Silvia Ferro}

\author[0000-0002-7526-8154]{Fabio Bacchini}
\affiliation{Centre for mathematical Plasma Astrophysics, Department of Mathematics, KU Leuven, Belgium}
\affiliation{Royal Belgian Institute for Space Aeronomy, Uccle, Belgium}
\email{fabio.bacchini@kuleuven.be}

\author[0000-0001-7233-2555]{Giuseppe Arrò}
\affiliation{Department of Physics, University of Wisconsin-Madison, USA}
\email{peppe.arro94@gmail.com}

\author[0000-0002-5272-5404]{Francesco Pucci}
\affiliation{CNR-ISTP, Bari, Italy}
\email{francesco.pucci@istp.cnr.it}

\author[0000-0002-6830-3767]{Pierre Henri}
\affiliation{Laboratoire Lagrange, Observatoire Côte d'Azur, CNRS, France}
\affiliation{LPC2E, CNRS, Université d'Orléans, France}
\email{pierre.henri@oca.eu}

\begin{abstract}
The Kelvin--Helmholtz instability (KHI) is a major driver of multiscale plasma dynamics at velocity shear layers, where it can promote the formation of current sheets and the onset of magnetic reconnection as well as drive plasma energization. While recent kinetic studies have shown efficient electron heating during nonlinear KH evolution, the connection between reconnection dynamics and localized electron energization is still not fully understood. 
We investigate this link using two-dimensional fully kinetic simulations of KHI developing in a double-periodic system with two velocity shear layers and a uniform guide field, initialized from a finite-Larmor-radius equilibrium. During the nonlinear stage, initially coherent vortices evolve into layers populated by fragmented current sheets \rev{containing active magnetic reconnection sites}.
The global energetics reveal \rev{distinct} species-dependent energization pathways. Ions act as the primary energy reservoir, transferring energy to the electromagnetic fields, while electrons receive the dominant net positive energy input. Electron energization is strongly anisotropic ($T_{\parallel,e} > T_{\perp,e}$) and localized within intermittent \rev{reconnecting} current sheets associated with enhanced field--particle energy exchange and elevated agyrotropy. These regions also show the development of suprathermal tails in the electron energy distributions, providing evidence for nonthermal electron energization. Despite opposite vorticity orientations, the two shear layers exhibit similar statistical behavior. Together, these results establish a direct connection between reconnection-associated current structures and localized electron energization in collisionless KHI dynamics.
\end{abstract}


\section{Introduction}
The Kelvin--Helmholtz instability (KHI) is a fundamental shear-driven process that develops at velocity interfaces in fluids, including magnetized plasmas, and plays an important role in a wide range of space and astrophysical environments, including planetary magnetopauses, heliospheric boundaries, and other plasma shear layers. At Earth’s low-latitude magnetopause, KH waves and rolled-up vortices have been frequently observed during intervals of northward interplanetary magnetic field, when magnetic reconnection is comparatively reduced and shear-driven growth is favored \citep{hasegawa2004, eriksson2016, stawarz2016, settino2024}. Similar signatures have also been reported at several other planetary magnetospheres, including Mercury, Mars, Saturn, and Jupiter \citep{sundberg2012,ma2015,ruhunusiri2016,ranquist2019,aizawa2020b,montgomery2023}.

The nonlinear evolution of the KHI can drive plasma transport, generate vortical and turbulent-like structures, distort magnetic fields, and promote the formation of thin current layers and associated magnetic-reconnection sites \citep{faganello2017}, which are expected to play a key role in plasma energization. Large-scale hydrodynamic and magnetohydrodynamic (MHD) simulations have shown that KH activity can contribute to boundary-layer transport, vortex evolution, and reconnection-mediated plasma entry at planetary magnetopauses and heliospheric boundaries \citep{faganello2012a, nakamura2013, ma2017, nakamura2017, ferro2024, ma2025}. Observations have also shown that KH activity can persist simultaneously on both terrestrial flanks, while local magnetic-shear and boundary deformation may influence its subsequent nonlinear evolution \citep{lu2019}.

Although the KHI has been extensively studied within MHD, many of the relevant plasma environments are weakly collisional or collisionless, so that isotropic-fluid descriptions become insufficient. In these regimes, pressure anisotropy, finite-Larmor-radius (FLR) effects, and full pressure-tensor dynamics can all influence both the onset and the nonlinear evolution of the instability. Recent work has demonstrated the significance of these kinetic effects: a study of the KHI in the CGL framework showed that anisotropic pressure can substantially modify the instability by diverting part of the available energy into parallel- and perpendicular-pressure channels, thereby reducing magnetic bending, reconnection activity, and intermittency relative to the MHD limit \citep{biswas2026}.

Kinetic studies have further shown that the relative orientation between the equilibrium flow vorticity and the magnetic field introduces an intrinsic asymmetry between opposite shear layers. In particular, comparative simulations across fluid, hybrid, and fully kinetic models demonstrated that this asymmetry is captured only when kinetic ion physics is retained, and that it affects the linear growth and early evolution of the instability \citep{henri2013}. Complementary extended-fluid equilibrium analyses have shown that FLR corrections can already imprint an asymmetry linked to the scalar product $\bb{\omega}\bcdot\bb{b}$, where $\bb{\omega}=\grad\btimes\bb{u}$ is the flow vorticity and $\bb{b}=\bb{B}/|\bb{B}|$ is the unit vector along the magnetic field. This modifies the current structure and can generate agyrotropic pressure configurations across magnetized shear layers \citep{cerri2018}. These results motivate the use of carefully constructed kinetic equilibria and indicate that shear layers with opposite $\bb{\omega}\bcdot\bb{b}$ may evolve differently even when the global-system geometry appears symmetric.

Beyond the question of instability growth, an important open issue is how energy is converted and deposited during the nonlinear evolution of collisionless shear flows. In collisionless plasmas, energization is often highly intermittent, with coherent current sheets and other localized structures concentrating field--particle energy exchange and plasma heating in specific regions rather than distributing it uniformly in space. Fully kinetic simulations have shown that such structures can extend down to the electron scales and are closely associated with enhanced heating and dissipation proxies \citep{wan2012,karimabadi2013}. Recent studies of shear-driven collisionless turbulence have further quantified species-dependent energy conversion and pressure--strain dissipation channels in such systems \citep{goodwill2025}.
In the specific context of the KHI, recent fully kinetic simulations have shown that the electron thermal energy can increase during the nonlinear evolution of the instability and have examined the associated energization channels using guiding-center diagnostics \citep{yang2026}.
Those results mainly reported an increase in electron temperature and identified candidate energization channels, but the relative importance of the underlying mechanisms remains debated. In particular, whether reconnection-associated current structures from nonlinear KH dynamics could drive an electron energy increase has not been quantitatively established.
More generally, the full collisionless-KHI energy-conversion pathway over sufficiently large ranges (from MHD down to electron scales) remains unresolved: it is still unclear how shear-flow free energy is redistributed between bulk motion, fields, and particles, which species dominates the net energization, and whether electron energization is predominantly thermal or also accompanied by non-Maxwellian features localized within intermittent current structures.

To address these questions, we present large-scale, two-dimensional (2D), fully kinetic particle-in-cell (PIC) simulations of a double-shear KHI in a configuration with a dominant out-of-plane guide field, using the same FLR-equilibrium framework adopted in our previous study \citep{ferro2026}. The double-shear setup enables an internal comparison between two simultaneously evolving layers with opposite vorticity orientations relative to the guide field. We analyze the system using complementary diagnostics, including the global energy budget, species-resolved $\bb{J}\bcdot\bb{E}$ (with $\bb{J}$ the current density and $\bb{E}$ the electric field) in the parallel and perpendicular directions, temperature anisotropy, and the agyrotropy measure $Q$ introduced by \cite{swisdak2016}. These diagnostics allow us to quantify the transfer of free energy between bulk flow, electromagnetic fields, and particle populations, and to directly link localized electron energization to reconnection-associated current structures formed during the nonlinear evolution of the instability.

The remainder of this Letter is organized as follows: \autoref{sec:method} briefly describes the numerical setup and diagnostics. \autoref{sec:results} presents the evolution of the instability, the global energy-conversion channels, and the link between reconnection-associated structures and localized electron energization. \autoref{sec:discussion} discusses the implications of these results and the comparison between the two shear layers. Finally, \autoref{sec:conclusions} summarizes our main conclusions.

\section{Methods}
\label{sec:method}
We perform 2D PIC simulations of the collisionless KHI using the same numerical setup as in \citet{ferro2026}. The simulations are carried out with the semi-implicit PIC code iPIC3D, implementing energy-conserving PIC algorithms \citep{markidis2010,lapenta2017a,bacchini2023}. The numerical parameters, normalization, and boundary conditions are described in \citet{ferro2026}; here we summarize only the elements relevant to the present analysis.

The system consists of a collisionless magnetized plasma with a double shear flow in the $x$-direction and a magnetic field with an in-plane ($B_{x,0}$) and an out-of-plane guide component ($B_{z,0}=10\,B_{x,0}$). The initial plasma conditions follow the finite-Larmor-radius (FLR) equilibrium of \citet{cerri2013}, which reduces artificial kinetic effects on the initial state and captures the dependence of the relative orientation between vorticity and magnetic field \citep{henri2013}. The two shear layers, hereafter referred to as the lower shear (LS) and upper shear (US), evolve simultaneously under opposite vorticity orientations relative to the guide field. We initialize an ion--electron plasma with mass ratio $64$ in a domain of $150\times400\,d_i^2$\rev{, discretized with $2304\times6144$ grid cells ($\Delta x=\Delta y=0.065\,d_i$)}, and run the simulation for over $1000\,\Omega_{c,i}^{-1}$, with $d_i$ the ion skin depth and $\Omega_{c,i}$ the ion cyclotron frequency.
\rev{The initial velocity profile has a shear half-width $\delta=3\,d_i$. Using the drift speed $V_{\rm drift}=0.028\,c$ as the characteristic flow speed, the initial sonic Mach number is $M_s=V_{\rm drift}/c_s=0.71$. The corresponding Alfv\'enic Mach numbers based on the magnetic-field components parallel and perpendicular to the flow are
$M_{A,\parallel}=V_{\rm drift}/v_{A,\parallel}=5.01$
and
$M_{A,\perp}=V_{\rm drift}/v_{A,\perp}=0.50$,
where
$v_{A,\parallel}=B_{\parallel}/\sqrt{4\pi\rho}$
and
$v_{A,\perp}=B_{\perp}/\sqrt{4\pi\rho}$.}

We focus our analysis on the nonlinear stage of the instability, when vortex interaction, the formation of thin current sheets and associated reconnection activity, electron energization, and kinetic departures from gyrotropy become most pronounced. Plasma energization and its association with reconnection-related structures are quantified using complementary global and local diagnostics. We decompose the total energy into electromagnetic, bulk, and thermal contributions for each species, and evaluate the field--particle energy-exchange rate $\bb{J}\bcdot\bb{E}$ separately for electrons and ions. This term is further decomposed into components parallel and perpendicular to the local magnetic field. We also compute the pressure--strain interaction and the compressional term to characterize bulk-to-thermal energy conversion. From the reconstructed pressure tensor, we obtain the parallel and perpendicular temperatures, $T_\parallel$ and $T_\perp$, and define the temperature anisotropy as $A = T_\perp/T_\parallel-1$. 
\rev{Departure from gyrotropy is characterized using the invariant-based measure $Q$ introduced by \citet{swisdak2016}, defined as
\begin{equation}
Q = 1 - \frac{4\,I_2}{(I_1 - P_{\parallel})(I_1 + 3P_{\parallel})},
\end{equation}
where $I_1=\mathrm{Tr}(\mathbf{P})$ and $I_2 = P_{xx}P_{yy}+P_{xx}P_{zz}+P_{yy}P_{zz}
-\left(P_{xy}^2+P_{xz}^2+P_{yz}^2\right)$
are the first and second invariants of the pressure tensor, and $P_{\parallel}$ is the pressure parallel to the local magnetic field. Throughout the paper, we use $\sqrt{Q}$ as a scalar proxy for pressure-tensor agyrotropy.}

\section{Results}
\label{sec:results}
\subsection{Global Collisionless-KHI Evolution}
\label{subsec:KHI_evolution}

\begin{figure}
    \centering
    \includegraphics[width=\linewidth]{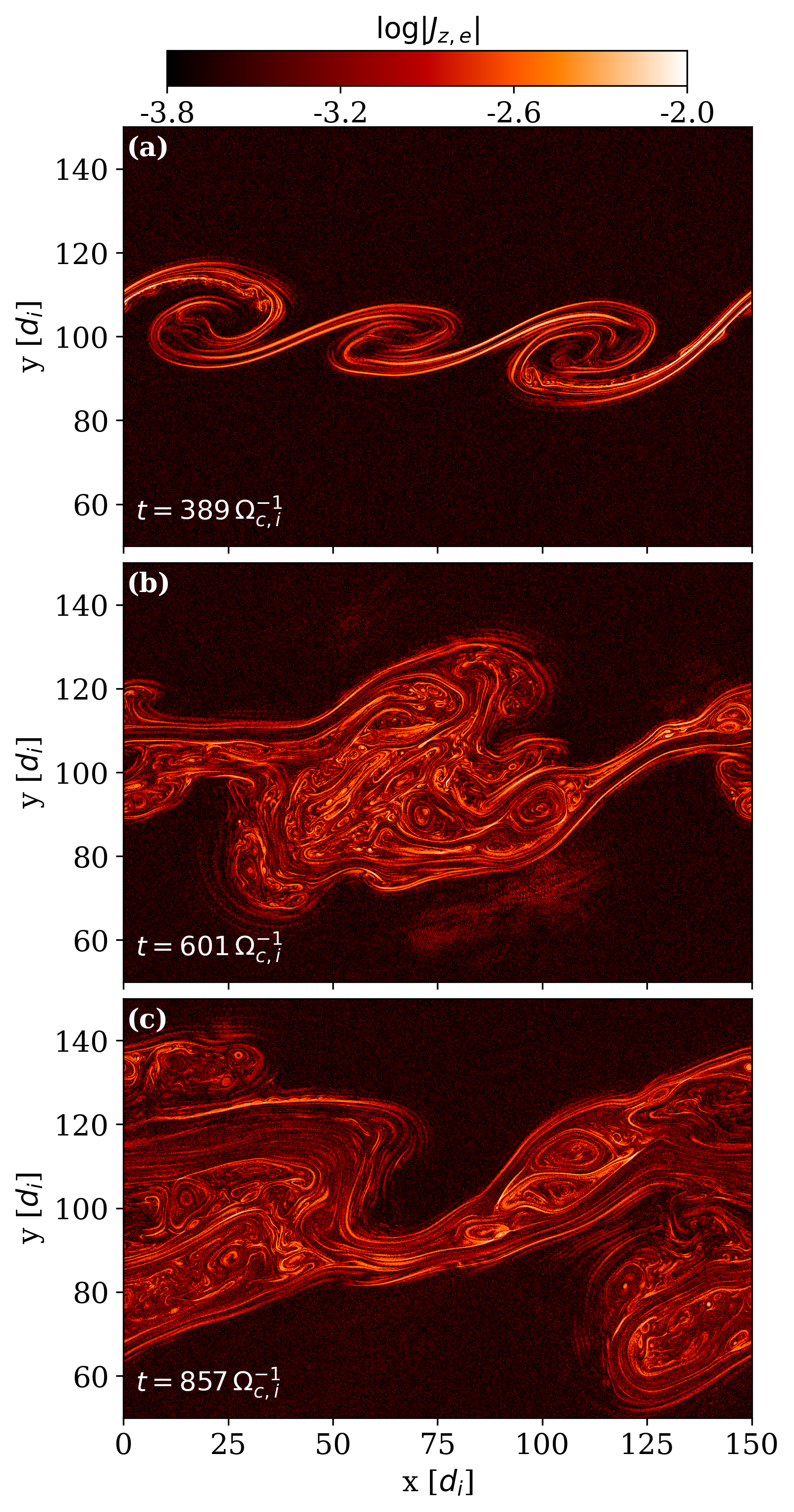}
    \caption{
    Temporal evolution of the electron out-of-plane current density in the LS, where the flow vorticity is parallel to the in-plane magnetic field. 
    We show representative stages of the instability: (a) early nonlinear phase, (b) vortex-merging phase, and (c) fully nonlinear turbulent phase.
    }
    \label{fig:Jz_evolution}
\end{figure}
The system evolves through the standard stages characteristic of the KHI: a linear phase ($t \lesssim 250\,\Omega_{c,i}^{-1}$), an early nonlinear phase with growing interacting vortices ($250 \lesssim t \lesssim 450\,\Omega_{c,i}^{-1}$), a vortex-merging phase ($450 \lesssim t \lesssim 650\,\Omega_{c,i}^{-1}$), and a late nonlinear turbulent-like stage ($t \gtrsim 650\,\Omega_{c,i}^{-1}$). The dominant mode initially leads to the formation of three vortices inside each shear layer.

During the nonlinear evolution, vortical advection and compression distort the magnetic field and generate localized current structures. As shown in \autoref{fig:Jz_evolution}, current layers first outline coherent vortices (panel (a)), then intensify during vortex merging (panel (b)), and finally fragment into thin intermittent filaments in the late nonlinear stage (panel (c)). This evolution is consistent with reconnection activity reported for the same configuration by \citet{ferro2026}. The presence of localized current sheets is central to the energization processes discussed below, as they are consistent with reconnection activity and coincide with enhanced $\bb{J}_e\bcdot\bb{E}$, strong electron heating, and departures from gyrotropic behavior.

\subsection{Global Energy Conversion}
\label{subsec:global_energy}
\begin{figure}
    \centering
    \includegraphics[width=\linewidth]{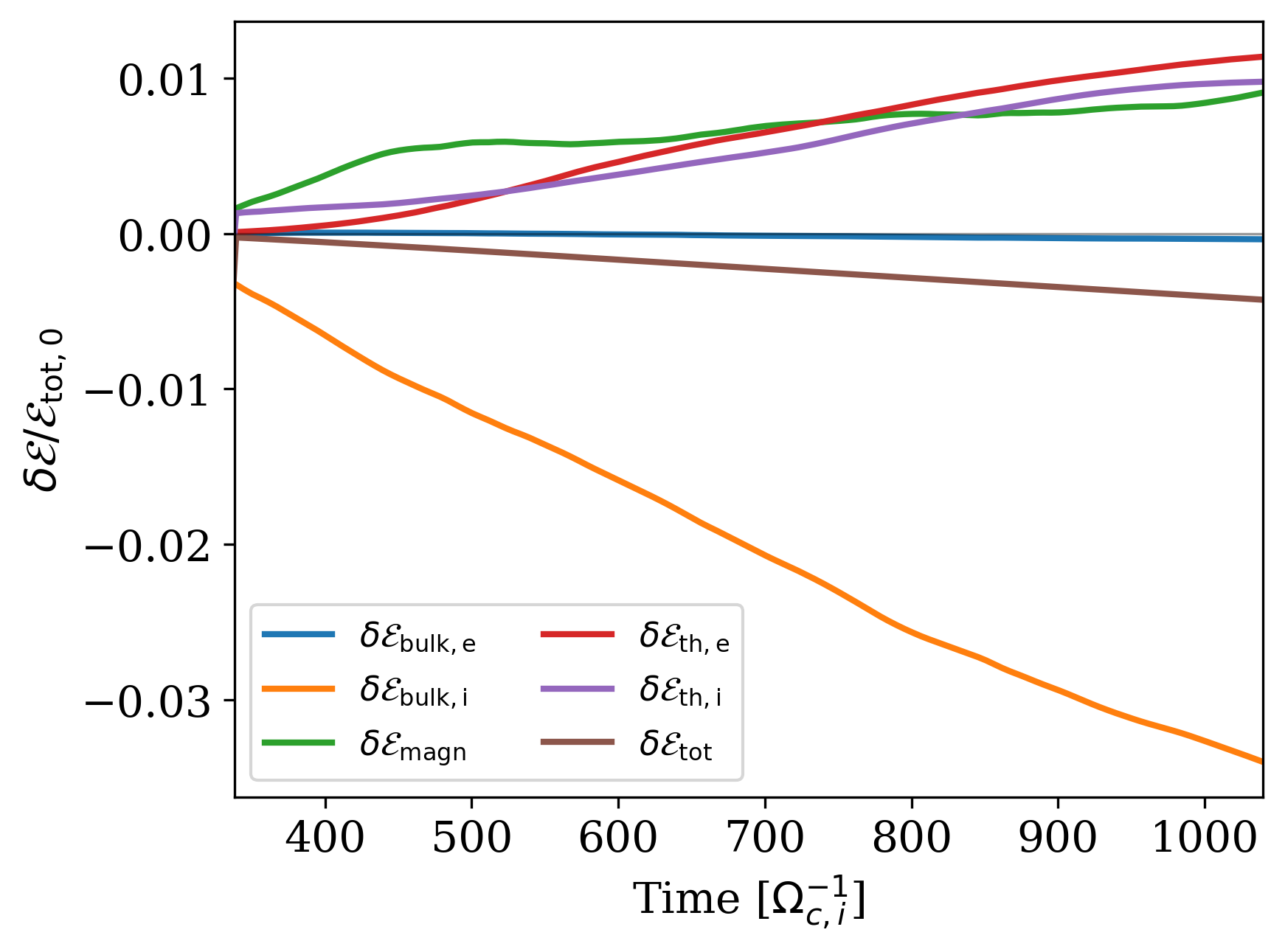}
    \caption{
    Temporal evolution of the difference in each domain-integrated energy component during the nonlinear KH stage, normalized to the initial total energy. 
    }
    \label{fig:tevol_energies1D}
\end{figure}

We examine the global energy budget during the nonlinear evolution of the KHI by decomposing the total energy into electromagnetic, bulk, and thermal contributions for each species. As the instability saturates, the bulk kinetic energy associated with the shear flow decreases while the electromagnetic energy increases. At later times, a significant fraction of this energy is converted into particle thermal energy, indicating the onset of sustained plasma heating (\autoref{fig:tevol_energies1D}).

\begin{figure}
    \centering
    \includegraphics[width=\linewidth]{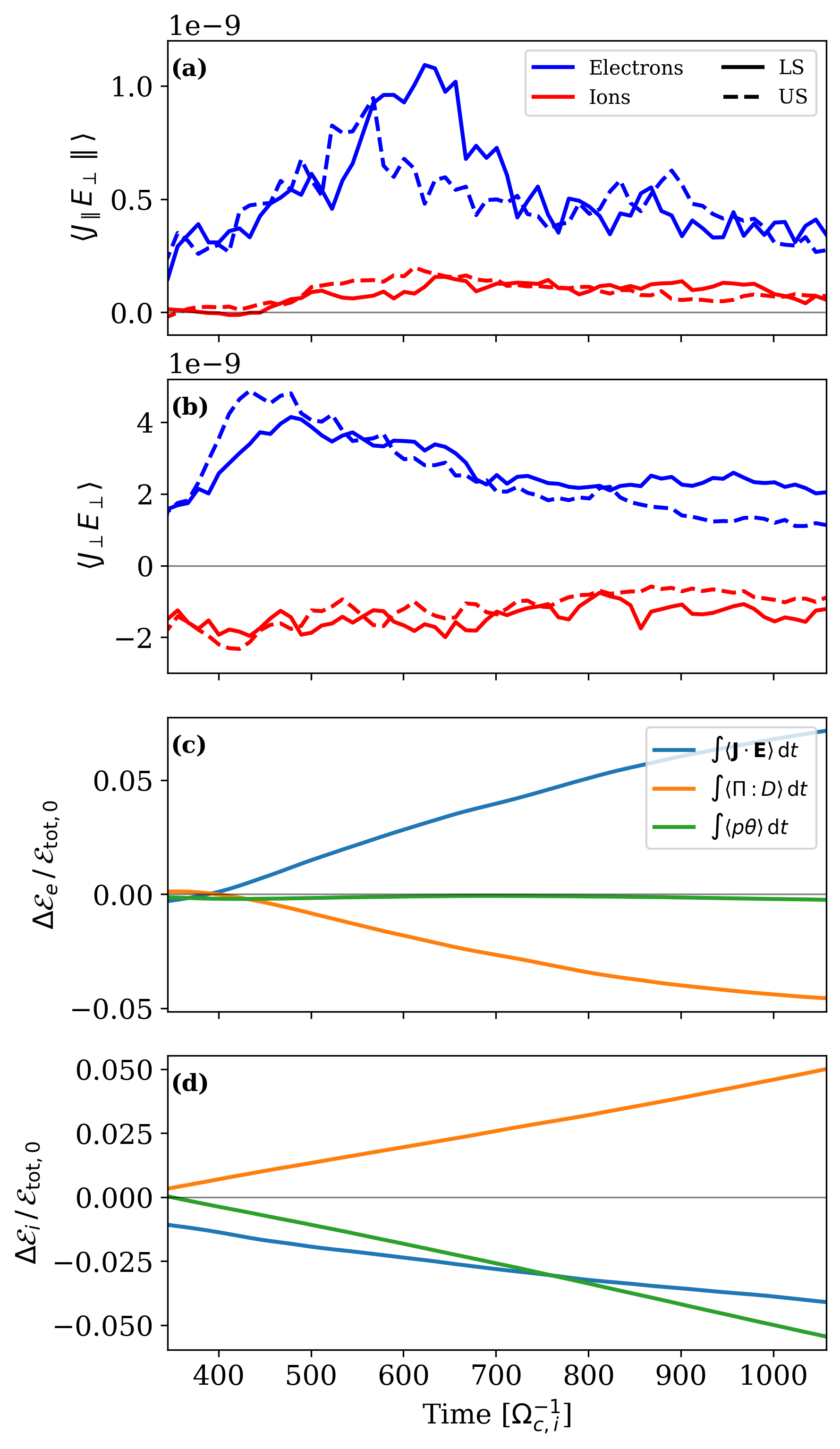}
    \caption{    
    Temporal evolution of the energy-transfer channels during the nonlinear stage of the KHI. 
    Panels (a) and (b) show the spatially averaged parallel and perpendicular components of the electromagnetic work, $\langle J_{\parallel}  E_{\parallel}\rangle$ and $\langle J_{\perp} E_{\perp}\rangle$, respectively, for electrons (blue) and ions (red), separately in the LS (solid lines) and US (dashed lines). 
    Panels (c) and (d) show the time-integrated cumulative contributions for electrons and ions, respectively: $\int \bb{J}\,\bcdot\,\bb{E}\,\rmd t$, $\int \Pi:D\,\rmd t$, and $\int p\theta\,\rmd t$, normalized to the initial total energy $\mathrm{E}_{\mathrm{tot},0}$.}
    \label{fig:energy_channels_pathways}
\end{figure}

To quantify the energy exchange between fields and particles, we analyze the energy-exchange channels available to collisionless plasmas (e.g.\ \citealt{yang2017b,matthaeus2020pathways,arro2022spectral}). Of particular importance is the work term $\bb{J}\bcdot\bb{E}$ for ions and electrons, which we decompose into parallel and perpendicular components relative to the local magnetic field. 
The term $\bb{J}\bcdot\bb{E}>0$ corresponds to an increase in plasma energy and an equal loss of electromagnetic energy. \autoref{fig:energy_channels_pathways}(a,b) shows that $\bb{J}_e\bcdot\bb{E}>0$ in both shear layers and is dominated by the perpendicular contribution. In contrast, ions exhibit a weaker response, with positive parallel and negative perpendicular contributions, indicating a net transfer of energy from ions to the fields. This behavior is similar in both shear layers. The dominance of the perpendicular component reflects the largely in-plane dynamics of the KHI in the presence of a strong guide field, whereas the smaller parallel contribution is more directly associated with nonideal field--particle interactions within localized current structures. 
As shown in the following subsection, these regions coincide with intense current sheets, enhanced parallel electron energization, and elevated agyrotropy, identifying them as localized sites of electron energy increase consistent with reconnection-related processes.

The global energetics and $\bb{J}\bcdot\bb{E}$ diagnostics \rev{indicate a coherent energy pathway in which ion bulk-flow energy is first transferred to the electromagnetic fields and subsequently deposited into electrons through both thermal and nonthermal energization.}

This interpretation is supported by the cumulative contributions shown in \autoref{fig:energy_channels_pathways}(c,d), where electromagnetic work provides the dominant positive net input for electrons and the dominant negative contribution for ions.
\rev{\rev{The conversion between bulk-flow and thermal energy can be decomposed into a compressive term $p\,\theta$ and a traceless pressure--strain interaction $\bb{\Pi}\bdbldot\textbf{D}$, where $p=\mathrm{Tr}(\mathbf{P})/3$ is the scalar pressure, $\mathbf{P}$ is the pressure tensor, $\theta=\grad\bcdot\bb{u}$ is the bulk-flow divergence, $\bb{u}$ is the bulk velocity, $\Pi_{ij}=P_{ij}-p\,\delta_{ij}$ is the traceless pressure tensor, and $\mathrm{D}_{ij}=(\partial_i u_j+\partial_j u_i)/2-\theta\,\delta_{ij}/3$ is the traceless strain-rate tensor.} Following the standard sign convention \citep{yang2017b,matthaeus2020pathways,arro2022spectral}, the thermal energy equation contains the combination $-(p\,\theta+\bb{\Pi}\bdbldot\textbf{D})$, so negative values of either term correspond to net thermal energization. The pressure--strain interaction exhibits different behavior for the two species. For electrons, $\bb{\Pi}\bdbldot\textbf{D}<0$, corresponding to net bulk-to-thermal energy conversion, while $p\,\theta$ remains close to zero. For ions, $\bb{\Pi}\bdbldot\textbf{D}>0$, indicating a transfer from thermal to bulk energy, whereas the negative values of $p\,\theta$ indicate compressive heating.}
Overall, these results indicate that the instability ultimately channels energy into localized electron energization within intermittent current structures, consistent with previous studies of kinetic energy conversion in magnetized plasmas \citep[e.g.][]{wan2012,karimabadi2013}. Interestingly, qualitatively similar energy-transfer pathways, in which ion bulk-flow energy provides the primary source for electron heating, have also been reported in other processes such as turbulence \citep{arro2022spectral} and phase mixing of plasma waves \citep{bacchini2022}, suggesting that this pathway to kinetic-scale heating is a \rev{rather general feature of collisionless} plasmas.

\subsection{Electron Energization at Magnetic Reconnection Sites}
\label{subsec:elec_heating}
\begin{figure}
    \centering
    \includegraphics[width=\linewidth]{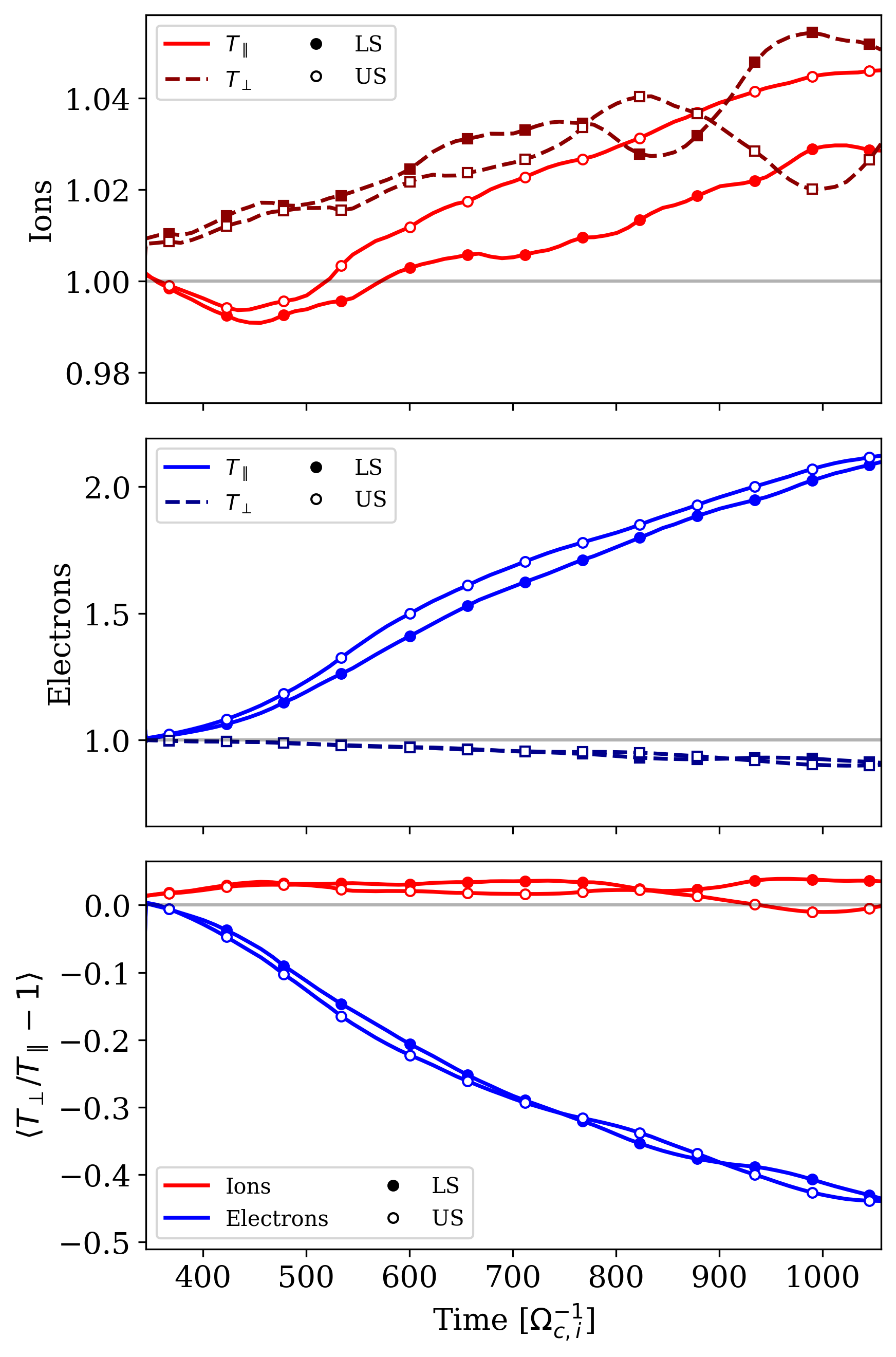}
    \caption{
    Temporal evolution of the average $T_\parallel$, $T_\perp$, and temperature anisotropy during the nonlinear phase in the two shear layers. Top: ions. Middle: electrons. Bottom: anisotropy.
    }
    \label{fig:tevol_Tanis}
\end{figure}

Here, we examine the temporal evolution of the parallel and perpendicular temperatures for electrons and ions during the nonlinear phase of the KHI. As shown in \autoref{fig:tevol_Tanis}, both species experience an overall temperature increase, consistent with the global energy conversion discussed above. However, the heating is strongly species-dependent: electrons experience a larger increase in temperature, while ions evolve more gradually. Electron heating is also highly anisotropic, with electrons developing a pronounced negative anisotropy ($T_{\parallel,e} > T_{\perp,e}$) that increases throughout the nonlinear stage, whereas ions remain comparatively closer to isotropy. This indicates that the energy transferred from the fields preferentially energizes electrons along the magnetic-field direction.

\begin{figure*}
    \centering
    \includegraphics[width=0.9\linewidth]{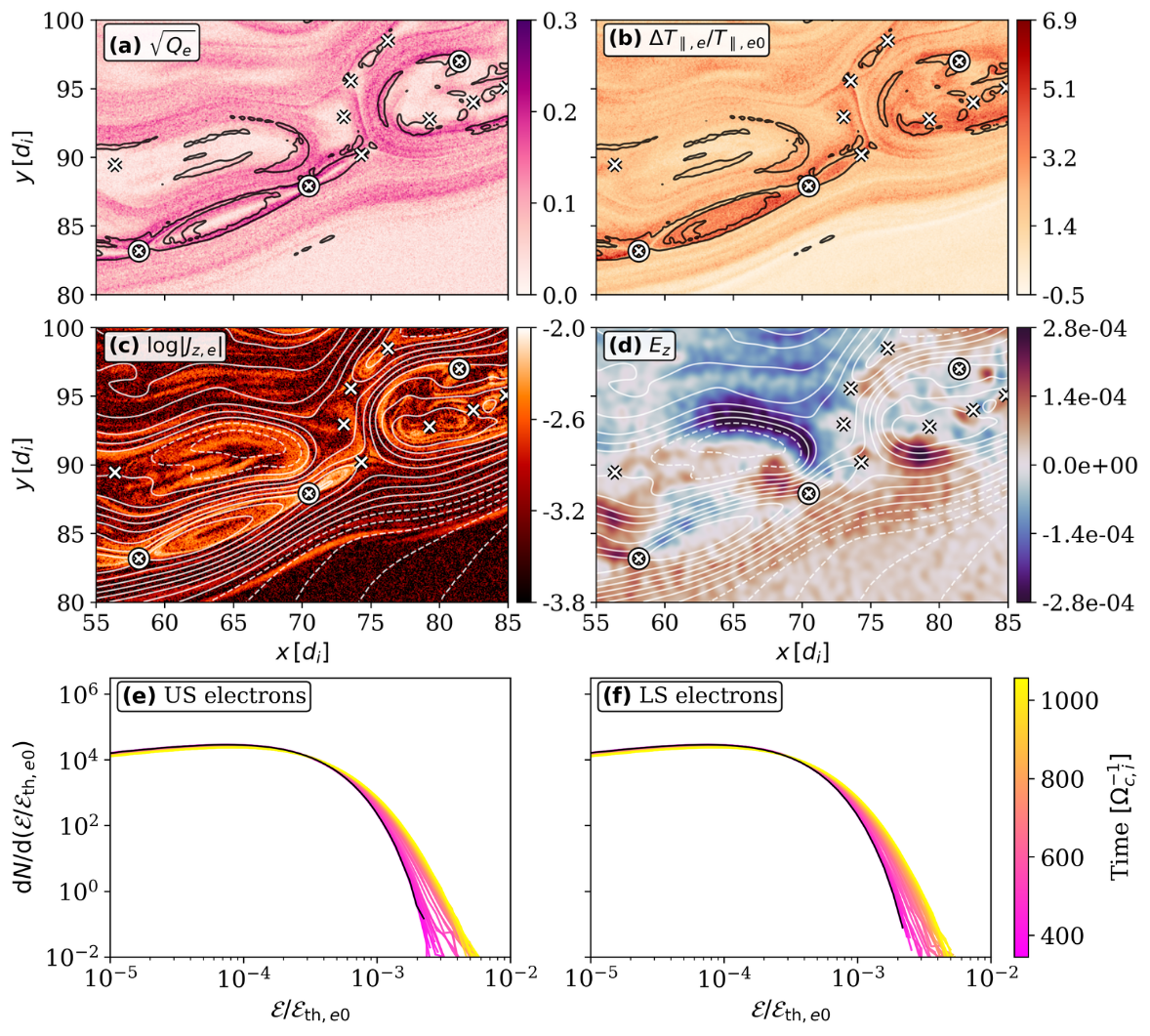}
    \caption{
    \rev{Top and middle rows: Zoom-in onto a $20 \times 30\,d_i^2$ region inside the lower shear layer (LS) during the late nonlinear stage of the KHI ($t=778\,\Omega_{c,i}^{-1}$).
    (a) Electron agyrotropy measure $\sqrt{Q_e}$.
    (b) Normalized parallel electron temperature variation, $\Delta T_{\parallel,e}/T_{\parallel,e0}$.
    (c) Logarithm of the out-of-plane electron current density, $\log |J_{z,e}|$.
    (d) Out-of-plane electric field, $E_z$.
    Black contours in panels (a,b) denote regions of intense electron current density, while white contours in panels (c,d) show magnetic-flux isocontours. White crosses indicate magnetic X-points, while circled X-points indicate the ones with the largest values of $|E_z|$, corresponding to the most active reconnection sites at the given time.
    Bottom row:
    (e,f) Temporal evolution of the normalized electron energy distributions in the upper (US) and lower (LS) shear layers, respectively. The black curve denotes the earliest distribution shown and serves as a reference for the progressive development of a suprathermal electron population during the nonlinear stage. Curves are colored according to time.}}
    \label{fig:heating_maps_spectra}
\end{figure*}

\rev{To identify where electron energization occurs within the shear layers, we examine the spatial distribution of agyrotropy, parallel electron heating, electron current density, the out-of-plane electric field, and the local magnetic topology} (\autoref{fig:heating_maps_spectra}a--\rev{d}). \rev{Strong enhancements of $\Delta T_{\parallel,e}/T_{\parallel,e,0}$ are concentrated within thin filamentary electron current sheets} embedded in the turbulent shear layer during the nonlinear KH stage. \rev{These regions also exhibit elevated values of the agyrotropy measure} $\sqrt{Q}$, indicating strong departures from gyrotropic-Maxwellian behavior.
\rev{The magnetic topology was reconstructed from the magnetic flux function $\psi$ (such that $\bb{B}=\grad\psi$), computed from the in-plane magnetic field, while magnetic X-points were identified automatically from the intersections of $B_x=0$ and $B_y=0$ contours and classified through the Hessian of $\psi$, following standard 2D reconnection diagnostics (\citealt{servidio2009,wan2013}). The reconstructed magnetic topology reveals numerous X-points embedded within the turbulent current-sheet network. However, only a subset of these X-points exhibits strong localized enhancements of the out-of-plane electric field $E_z$, commonly used as a proxy for the local reconnection rate. These active X-points coincide with the most intense electron current layers and with localized enhancements of parallel electron heating and agyrotropy, identifying them as the dominant electron energization sites. Representative examples are the X-points located near $(x,y)\approx(57,83)\,d_i$ and $(x,y)\approx(70,87)\,d_i$, which lie within a thin current sheet separating a chain of plasmoids. A detailed local analysis of one representative X-point (\autoref{app:local_reconnection}) demonstrates the characteristic signatures of collisionless magnetic reconnection, including the reversal of the reconnecting magnetic field across the current sheet, a localized electron current layer, a clear bidirectional electron exhaust, and a weaker ion response consistent with the developing reconnection outflow. Together, these diagnostics establish that the localized electron heating observed in \autoref{fig:heating_maps_spectra} is directly associated with ongoing magnetic reconnection.}
Electron energization is, therefore, highly intermittent rather than volume-filling, with energy deposition concentrated within a limited fraction of the turbulent domain. Combined with the local reconnection analysis presented above, the statistical correlation between current intensity and parallel electron heating discussed in \autoref{app:heating} indicates that the strongest heating events occur within reconnecting current sheets generated during the nonlinear evolution of the KHI.

To further characterize the nature of electron energization, we examine the evolution of the electron energy distributions in the two shear layers (\autoref{fig:heating_maps_spectra}(\rev{e,f})). Relative to the earliest distribution shown (black curve), electrons progressively develop a high-energy tail during the nonlinear stage, indicating the emergence of suprathermal and nonthermal features (see an example of a representative velocity distribution in \autoref{app:vdf}). This behavior is observed in both shear layers and is \rev{consistent with electron acceleration at the active reconnection sites identified in the spatial analysis, where enhanced $E_z$, intense electron currents, agyrotropy, and bidirectional plasma exhausts demonstrate ongoing magnetic reconnection.}

Overall, energy is transferred from large-scale shear flows to electromagnetic fluctuations and ultimately deposited into electrons within intermittent current sheets. Rather than occurring uniformly throughout the turbulent layer, the strongest electron energization is concentrated at active magnetic reconnection sites, where enhanced $E_z$, agyrotropy, parallel heating, localized electron currents, and bidirectional plasma exhausts coexist.

\section{Discussion}
\label{sec:discussion}

\subsection{Nature of Energy Dissipation in Collisionless KHI}
\label{subsec:nature_dissipation}
Our results show that energy dissipation in the collisionless KHI is strongly intermittent and mediated \rev{by localized reconnecting current sheets, where magnetic reconnection drives strong field--particle interactions and electron energization}. During the nonlinear evolution, initially coherent vortex-scale current layers progressively fragment into thin filaments and localized gradients (\autoref{fig:Jz_evolution}), indicating a transfer of energy from large fluid scales toward ion and electron kinetic scales. Particle energization is concentrated within these structures rather than distributed throughout the full system, consistent with previous studies of collisionless plasma turbulence that identified intermittent current sheets as preferred sites of dissipation and energy conversion \citep{wan2012,karimabadi2013}. This result applies under the specific conditions we chose for this study, which are, however, highly relevant for magnetospheric plasmas.

The behavior we observe contrasts with commonly employed fluid-based descriptions, where heating is typically represented through effective viscous or resistive dissipation, whereas in weakly collisional plasmas energy conversion proceeds through kinetic processes concentrated in coherent structures and field--particle interactions \citep{matthaeus2015,howes2017,matthaeus2020pathways}. 
In this regime, dissipation is therefore localized and acts over a limited fraction of the domain. The spatial correspondence of intense current density, enhanced parallel electron heating, and elevated agyrotropy indicates that the most active energy-conversion sites are also the regions where fluid closures break down most strongly.
These findings place the collisionless KHI within the broader class of shear-driven kinetic plasma systems in which turbulence transfers energy to small scales before dissipation proceeds through coherent structures. Previous studies of collisionless turbulence and shear-driven kinetic systems have similarly reported intermittent dissipation, pressure--strain-mediated heating, and species-dependent energy conversion \citep{servidio2012local,wan2012,karimabadi2013,goodwill2025}. 
Recent observations of KH vortices from the Magnetospheric Multiscale (MMS) mission further support this interpretation, showing that energy transfer within vortex regions is highly intermittent and associated with enhanced kinetic activity and local nonthermal features \citep{settino2026}. This observational picture is broadly consistent with our results and supports the idea that KHI dissipation is intrinsically multiscale, intermittent, and sensitive to the nonlinear state of the vortices. In particular, the MMS analysis of \citet{settino2026} suggests that the dominant energy-conversion channels evolve with vortex development, with early-stage vortices characterized by net conversion of internal energy into bulk plasma motion, whereas more rolled-up vortices exhibit stronger bulk-to-thermal energy transfer.
\rev{Here we show that, in the specific KHI context, these processes are closely associated with vortex growth, vortex merging, magnetic-field distortion, and the formation of KH-generated current sheets. Local analysis demonstrates that the strongest electron heating occurs at active reconnection sites embedded within these current sheets (\autoref{app:local_reconnection}). This interpretation is consistent with previous studies showing that plasma turbulence naturally generates intermittent reconnecting current sheets at electron scales \citep{califano2020,arro2020}.} Our results further suggest that electron energization during the nonlinear vortex-merging stage is not limited to a simple thermal increase. In contrast to earlier kinetic studies that primarily reported temperature enhancements \citep{yang2026}, the present larger-domain simulation reveals localized non-Maxwellian features in regions of intense current activity \rev{(Appendix~\ref{app:vdf})}, indicating that electron energization there is not purely thermal but includes nonthermal components. This indicates that vortex coalescence can drive richer phase-space dynamics than captured by fluid moments alone and may represent an additional pathway toward efficient particle energization in collisionless shear flows.

The species-resolved energetics reveal a clear separation of roles between ions and electrons. Ions act primarily as the initial energy reservoir, supplying energy to the fields through the gradual reduction of bulk-flow energy. Electrons, in contrast, receive a substantial fraction of this transferred energy and undergo the strongest energization response.
This asymmetry is visible in both the instantaneous and cumulative diagnostics (\autoref{fig:energy_channels_pathways}). The ion perpendicular work term is predominantly negative, indicating that ions feed electromagnetic fluctuations, whereas the electron contribution remains positive in both channels. Integrated over time, the cumulative electromagnetic work is the dominant positive contribution for electrons and the dominant negative contribution for ions. A consistent energy pathway, therefore, emerges in which ion bulk energy is first converted into electromagnetic fluctuations and subsequently deposited into electrons through localized kinetic processes within current sheets.
Electron energization is not only stronger in magnitude but also qualitatively different from the ion response. Electrons develop sustained $T_{\parallel,e}>T_{\perp,e}$ together with elevated agyrotropy, whereas ions remain comparatively closer to isotropy and exhibit weaker kinetic signatures. This suggests that electrons are the primary recipients of the late-stage dissipation produced by nonlinear KHI turbulence.
More broadly, these results indicate that collisionless KHI energy transfer cannot be described as a simple single-fluid conversion of bulk energy into isotropic temperature. Instead, the energy partition is species-dependent, anisotropic, and mediated by localized kinetic processes associated with intermittent current sheets that serve as the primary sites of electron energization.

\subsection{Symmetry Between the Two Shear Layers}
\label{subsec:shearlayers_symmetry}

Our setup contains two shear layers with opposite vorticity, providing an internal comparison where both layers evolve simultaneously under identical global plasma parameters. Despite local morphological differences, the two shear layers exhibit remarkably similar statistical behavior. This contrasts with earlier fully kinetic studies, which reported different growth rates and nonlinear evolution between opposite shear layers due to the sign of $\bb{B}\bcdot\bb{\Omega}$ and ion FLR effects \citep{henri2013,nakamura2010}. In the present case, the similar behavior of the two layers may \rev{instead reflect the use of a refined kinetic equilibrium, which minimizes the early readjustment of the shear layers. As shown by \citet{nakamura2010}, in shifted-Maxwellian initializations the effective post-readjustment shear width can depend on the sign of $\bb{B}\bcdot\bb{\Omega}$, and the subsequent KH growth depends primarily on this final width. }
In both regions, we observe comparable growth and fragmentation of current structures, analogous electron heating trends, and consistent agyrotropic signatures. The temporal evolution of the temperature anisotropy and current activity is likewise similar in the two layers, particularly during the nonlinear stage. This correspondence strengthens the interpretation of the heating mechanisms identified here. Because the same signatures arise in two independently evolving shear layers, the observed electron energization cannot be attributed to a single transient structure or local isolated event, but instead reflects a robust property of the nonlinear KHI under symmetric conditions. 

At the same time, this idealized configuration differs from planetary magnetopauses, where the two flanks are generally not equivalent. At Earth, the dawn and dusk magnetopause can exhibit systematic asymmetries related to the orientation of the interplanetary magnetic field, bow-shock geometry, and unequal turbulence levels in the magnetosheath. Observations have reported both asymmetric occurrence rates of Kelvin--Helmholtz waves and intervals of quasi-symmetric propagation, indicating that the degree of dawn--dusk asymmetry depends on upstream conditions and local geometry \citep{lu2019}. Recent statistics from the MMS further show enhanced dusk-side electron vorticity and stronger turbulent activity in parts of the magnetosheath \citep{li2024}.
Comparable or even stronger flank asymmetries are expected at Mercury, where the smaller magnetosphere and the larger ion kinetic scales relative to system size make FLR effects especially important. Hybrid simulations and MESSENGER observations have shown that KH vortices can preferentially develop on the dusk flank, while growth on the dawn side is reduced by kinetic broadening of the shear layer and by convection-electric-field effects \citep{paral2013}.

The symmetric double-shear system studied here should therefore be regarded as a \textit{controlled reference case}. By isolating the intrinsic kinetic-heating processes of the nonlinear KHI in two statistically comparable layers, it provides a useful baseline for interpreting more realistic asymmetric magnetopause configurations in future studies.

\section{Conclusions}
\label{sec:conclusions}

\rev{We investigated the nonlinear evolution of the collisionless KHI using fully kinetic two-dimensional simulations of a double-shear configuration with a uniform guide field, with the aim of identifying how and where electron energization occurs.}

\rev{The nonlinear KHI transfers energy from large-scale shear flows to ion and electron kinetic scales through fragmented current sheets. Global energetics reveal that ions transfer energy to electromagnetic fluctuations, which subsequently energize electrons, producing enhanced parallel heating, temperature anisotropy, and nonthermal populations.}
\rev{Investigating the magnetic topology and local reconnection diagnostics, we identify active magnetic reconnection sites as the locations of the strongest electron energization. Their spatial association with enhanced parallel heating, agyrotropy, and intense electron currents establishes magnetic reconnection as the dominant energization mechanism during the nonlinear stage of the KHI.}
\rev{The two shear layers exhibit remarkably similar statistical behavior despite their opposite vorticity orientations, indicating that this energization pathway is robust under otherwise symmetric plasma conditions.}

\rev{Overall, our results show that energy extracted from the shear flow is transferred through electromagnetic fluctuations and ultimately deposited into electrons at intermittent magnetic reconnection sites within the turbulent shear layer. These findings provide a framework for interpreting energy conversion and particle energization at planetary magnetopauses and other collisionless shear layers where the KHI develops, and are directly relevant to the analysis of in-situ measurements from missions such as MMS. Future studies of fully three-dimensional and asymmetric configurations will determine how robust these pathways remain under more realistic space-plasma conditions.}

\begin{acknowledgments}
S.F.\ is supported by the FWO PhD fellowship ``Investigating Magnetospheric Plasma Dynamics with Large--scale Fully Kinetic Simulations'' (grant no.\ 1126325N). S.F.\ thanks Oreste Pezzi for valuable discussions and insightful comments that helped improve this work.
F.B.\ acknowledges support from the FED-tWIN programme (profile Prf-2020-004, project ``ENERGY'') issued by BELSPO. 
F.B.\ and F.P.\ acknowledge support from the FWO Junior Research Project G020224N granted by the Research Foundation -- Flanders (FWO). 
H.P. is supported by CNES.
The resources and services used in this work were provided by the VSC (Flemish Supercomputer Center), funded by the Research Foundation - Flanders (FWO) and the Flemish Government.
\end{acknowledgments}

\begin{contribution}
S.F.\ carried out the simulations, performed the data analysis, and wrote the manuscript. F.B.\ closely supervised the work. All authors contributed to the interpretation of the results, participated in the scientific discussion, provided feedback on the analysis, and commented on the manuscript.
\end{contribution}

\software{\textsc{iPIC3D} (\citealt{markidis2010,lapenta2017a,bacchini2023,shukla2025})}

\bibliographystyle{aasjournal}
\bibliography{biblio}

\appendix

\section{Local signatures of magnetic reconnection}
\label{app:local_reconnection}

\rev{To verify that the magnetic X-points identified from the in-plane magnetic topology in \autoref{fig:heating_maps_spectra} correspond to active reconnection sites, we examine the local structure of a representative current sheet during three consecutive simulation outputs, at $t=773$, $779$, $784$, and $790$ $\Omega_{c,i}^{-1}$. The local coordinate system is determined from the Hessian of the magnetic flux function $\psi$ evaluated at the X-point \citep{espinoza2026secondary}. The direction tangent to the current sheet is identified with the local exhaust direction, $\hat{\bb{t}}$, while the perpendicular direction defines the inflow or normal direction, $\hat{\bb{n}}$.}

\begin{figure*}
    \centering
    \includegraphics[
        width=\textwidth
    ]{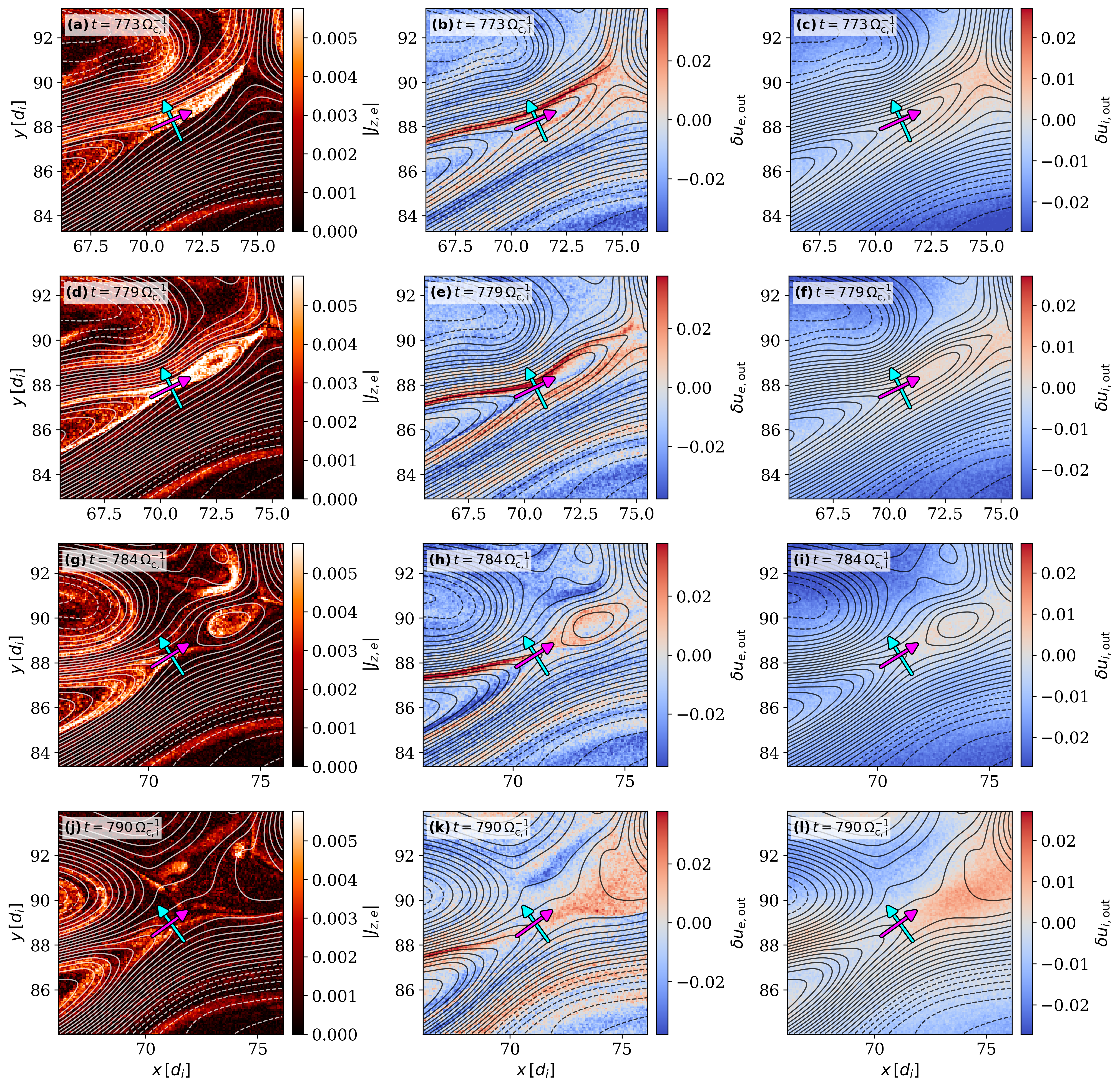}
    \caption{
    \rev{Local structure around a representative magnetic X-point at $t=773$, $779$, $784$, and $790~\Omega_{c,i}^{-1}$.
    The left column shows the electron out-of-plane current density $|J_{z,e}|$, while the middle and right columns show the projected electron and ion outflow velocities, $\delta u_{e,\mathrm{out}}$ and $\delta u_{i,\mathrm{out}}$, respectively, defined relative to the center-of-mass velocity at the X-point.
    Black contours represent the magnetic flux function $\psi$.
    The magenta triangle marks the X-point, while the magenta and cyan arrows indicate the local exhaust and normal directions, $\hat{\bb{t}}$ and $\hat{\bb{n}}$, respectively.}
    }
    \label{fig:local_reconnection_maps}
\end{figure*}

\rev{\autoref{fig:local_reconnection_maps} shows the electron out-of-plane current density, together with the electron and ion bulk velocities projected along the local exhaust direction. To highlight the velocity structure relative to the selected X-point, both electron and ion velocities are expressed in the local center-of-mass reference frame. Specifically, we subtract the center-of-mass velocity evaluated at the X-point,
\begin{equation}
    \delta u_{s,\mathrm{out}}
    =
    \left(
    \bb{u}_s
    -
    \bb{u}_{\mathrm{cm}}(\mathrm{XP})
    \right)
    \bcdot
    \hat{\bb{t}},
    \qquad
    s\in\{e,i\},
\end{equation}
to highlight the presence of reconnection jets, removing the background bulk velocity due to the combined motion of ions and electrons \citep{arro2023}. The black contours show the magnetic flux function $\psi$, and therefore trace the in-plane magnetic-field lines. The magenta and cyan arrows indicate the local exhaust and normal directions, respectively. A narrow electron current layer develops around the magnetic separatrix structure, while the projected electron and ion velocities display oppositely directed flows on the two sides of the X-point along the exhaust direction.}

\begin{figure*}
    \centering
    \includegraphics[
        width=\textwidth
    ]{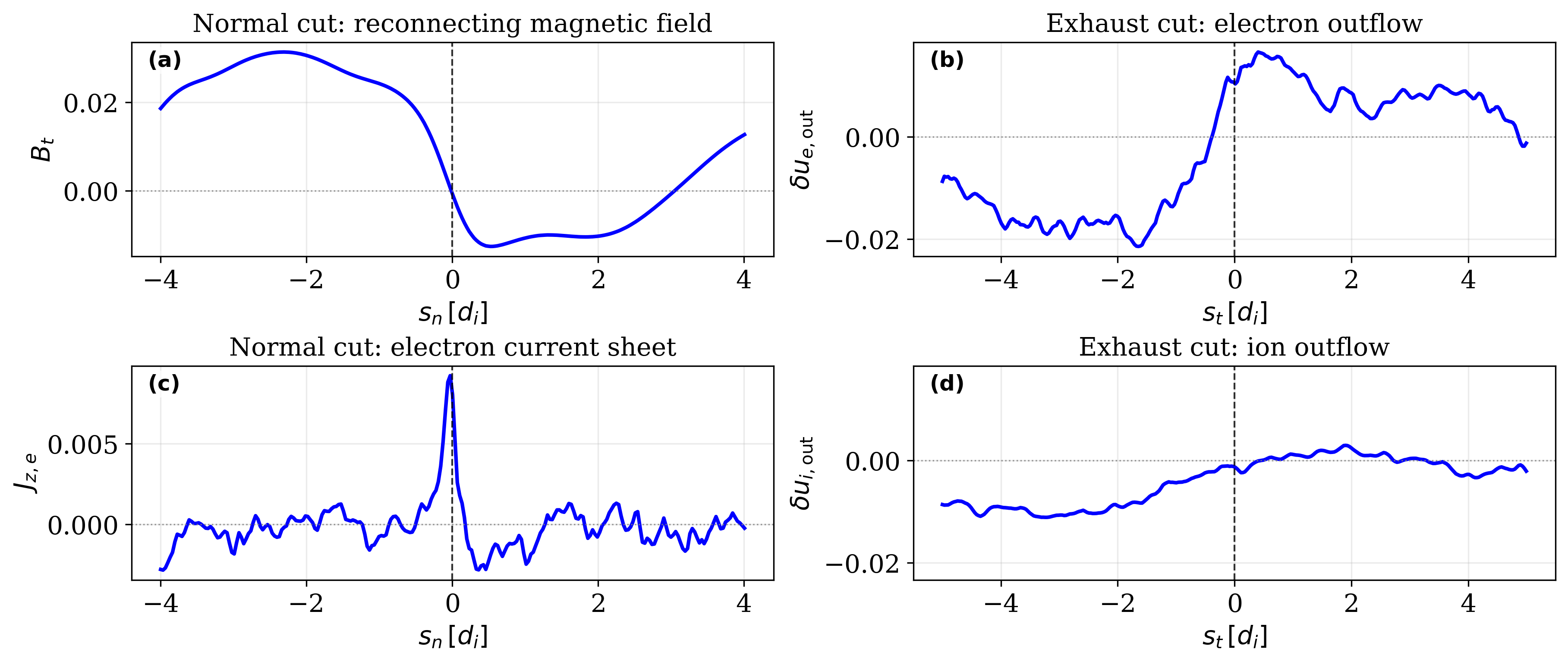}
    \caption{
    \rev{One-dimensional cuts through the X-point shown in \autoref{fig:local_reconnection_maps} at $t=784~\Omega_{c,i}^{-1}$.
    Panels (a) and (c) are taken along the local normal direction $\hat{\bb{n}}$ and show, respectively, the reconnecting magnetic-field component $B_t$ and the electron out-of-plane current density $J_{z,e}$.
    Panels (b) and (d) are taken along the local exhaust direction $\hat{\bb{t}}$ and show the projected electron and ion outflow velocity, $\delta u_{e,\mathrm{out}}$ and $\delta u_{i,\mathrm{out}}$, defined relative to the center-of-mass velocity at the X-point.
    The coordinate origin marks the X-point position.}
    }
    \label{fig:local_reconnection_profiles}
\end{figure*}

\rev{The corresponding one-dimensional profiles for the representative snapshot at $t=784~\Omega_{c,i}^{-1}$ are presented in \autoref{fig:local_reconnection_profiles}. Across the current sheet, the reconnecting magnetic-field component, $B_t=\bb{B}\bcdot\hat{\bb{t}}$, changes sign across the X-point, while $J_{z,e}$ exhibits a narrow localized peak. Along the exhaust direction, the projected electron and ion velocities display oppositely directed outflows on either side of the X-point. Although the profiles exhibit fluctuations associated with the surrounding turbulent environment, they retain the characteristic signatures of a reconnection exhaust, with a stronger and more structured electron response than ion response.}

\rev{The simultaneous presence of a magnetic-field reversal, a localized electron current sheet, an X-type magnetic topology, and a clear bidirectional electron exhaust, together with a comparatively weak ion response, provides direct local evidence that the selected structure is an active reconnection site. The pronounced asymmetry between the electron and ion exhausts suggests that reconnection may remain primarily electron-scale at this stage of the nonlinear evolution. A dedicated analysis of the ion-scale current-sheet structure and exhaust would, however, be required to determine whether the event satisfies the criteria for electron-only reconnection \citep[e.g.,][]{phan2018,sharma2019transition,arro2020,califano2020}.}

\section{Additional Diagnostics of Localized Electron Energization}
\label{app:heating}

\begin{figure}
    \centering
    \includegraphics[width=\linewidth]{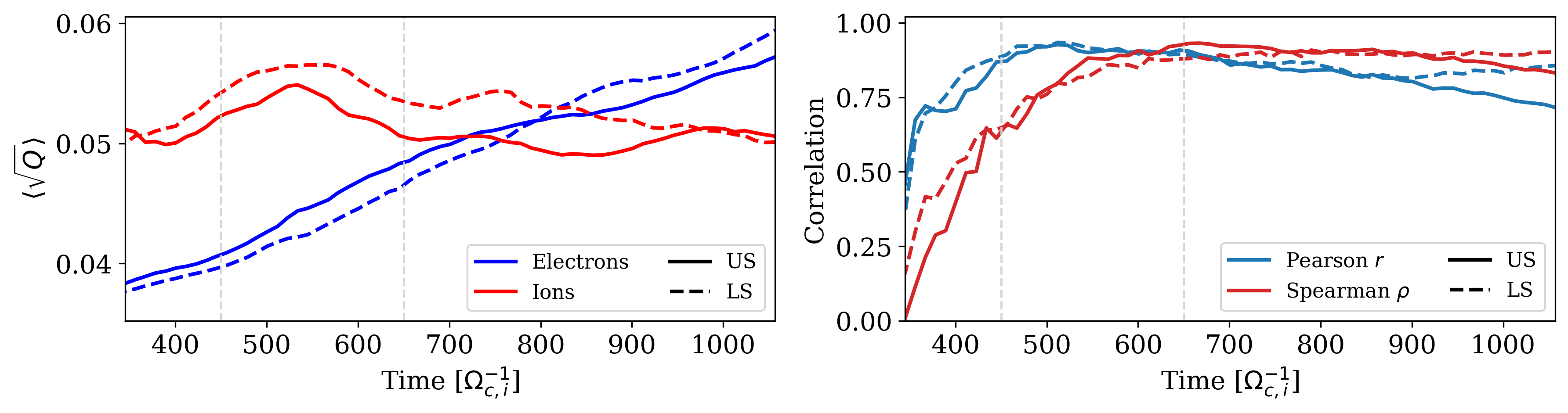}
    \caption{
    Additional diagnostics of localized electron energization. 
    Left: temporal evolution of the spatially averaged agyrotropy measure $\langle\sqrt{Q}\rangle$ for electrons and ions in the upper (US) and lower (LS) shear layers. 
    Right: Pearson ($r$) and Spearman ($\rho$) correlation coefficients between coarse-grained $\log |J_{z,e}|$ and $T_{\parallel,e}/T_{\parallel,e,0}$ in the same regions. 
    Vertical dashed lines mark the transitions between KHI stages.
    }
    \label{fig:app_corr_q}
\end{figure}

To further quantify the relation between intermittent current structures, electron energization, and kinetic departures from gyrotropy, \autoref{fig:app_corr_q} shows two complementary diagnostics for the upper and lower shear layers.

The left panel of \autoref{fig:app_corr_q} shows the time evolution of the spatially averaged agyrotropy measure $\langle \sqrt{Q} \, \rangle$ \citep{swisdak2016} for electrons and ions in both shear layers. In \autoref{fig:heating_maps_spectra}(a,b), enhanced local values of $\sqrt{Q}$ were shown to coincide with filamentary current structures and parallel heating sites. The temporal evolution in \autoref{fig:app_corr_q} confirms that these localized signatures are not isolated events but persist throughout the nonlinear stage. Electrons exhibit a steady increase of $\sqrt{Q}$ with time, reaching systematically larger values than ions, whereas ion agyrotropy remains comparatively weaker and more slowly varying. This species-dependent behavior is consistent with the preferential transfer of energy to electrons inferred from global energetics and with the stronger electron temperature anisotropy reported in \autoref{fig:tevol_Tanis}. The sustained growth of electron agyrotropy therefore reinforces the conclusion that late-stage KH-driven energization is intrinsically kinetic, occurring in regions where the electron distribution function departs significantly from gyrotropic Maxwellian equilibrium.

The right panel of \autoref{fig:app_corr_q} quantifies the statistical association between intense current structures and parallel electron heating through the Pearson linear correlation coefficient $r$ \citep{pearson1895} and the Spearman rank correlation coefficient $\rho$ \citep{spearman1904}. The Pearson coefficient measures the strength of a linear relation between two variables, whereas the Spearman coefficient evaluates whether the variables follow a monotonic trend independently of the exact functional form. Here, both diagnostics are computed from coarse-grained values of the out-of-plane electron current density, $\log |J_{z,e}|$, and the normalized parallel electron temperature, $T_{\parallel,e}/T_{\parallel,e,0}$, separately in the upper and lower shear layers. The coarse-graining procedure reduces cell-scale noise and tests whether regions of enhanced current also tend to exhibit stronger parallel electron energization. In both layers, $r$ and $\rho$ rapidly increase during the early nonlinear and merging phases, reaching values close to unity in the late nonlinear phase of the KHI (vertical dashed lines in \autoref{fig:app_corr_q}). The simultaneous enhancement of both coefficients demonstrates that regions of stronger current systematically correspond to larger parallel electron heating, and that this relation is not limited to a purely linear dependence. The lower shear reaches slightly higher peak values at intermediate times, while the upper shear remains somewhat more stable at later times, indicating only modest quantitative asymmetries. Overall, the strong correlation in both layers supports the interpretation advanced in \autoref{subsec:elec_heating} that intermittent current sheets are the preferred sites of electron energization.

\section{Example of a Localized Non-Maxwellian Electron Velocity Distribution Function}
\label{app:vdf}
\begin{figure*}
    \centering
    \includegraphics[width=\textwidth]{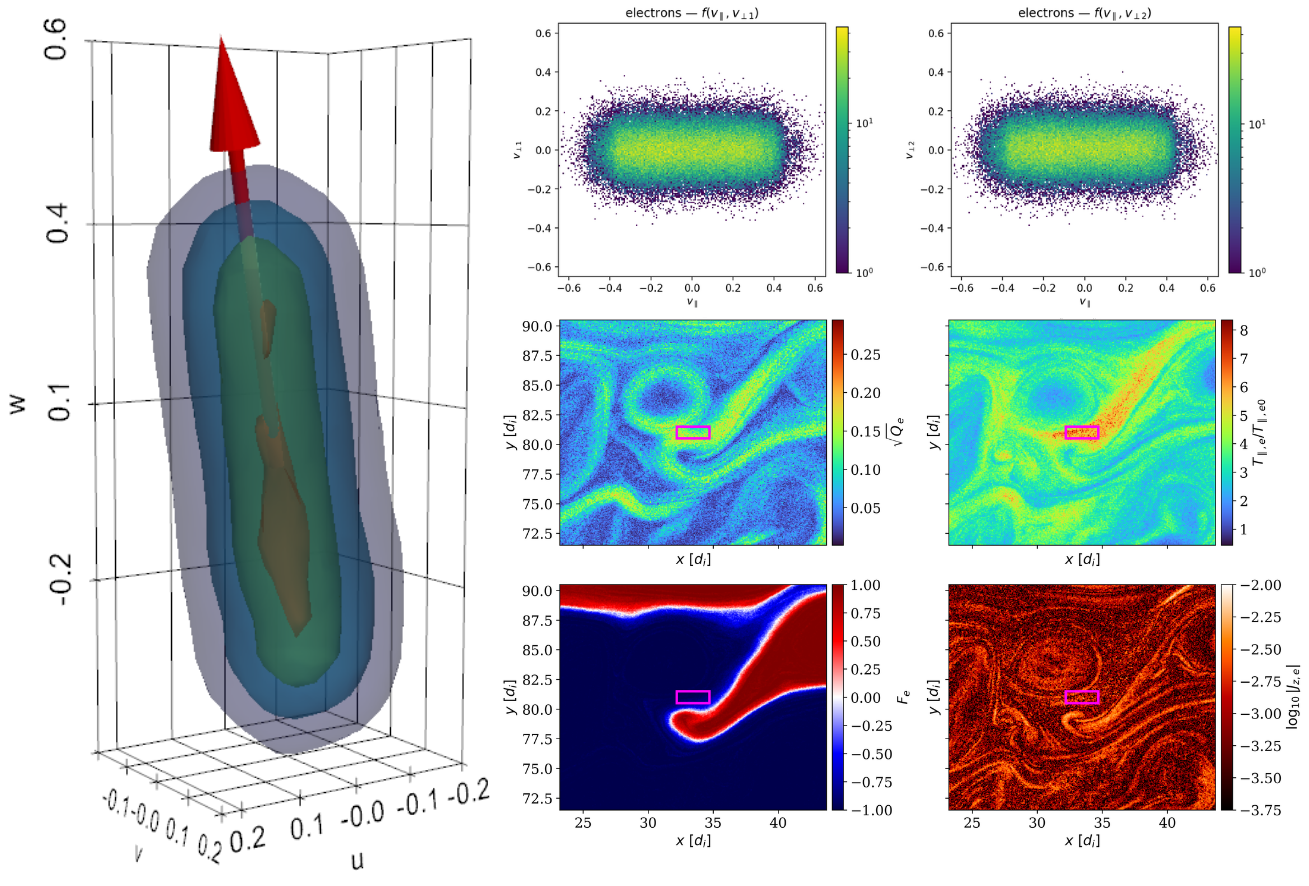}
    \caption{
    Example of a localized subregion in the lower shear layer during the nonlinear stage. 
    Left: Three-dimensional visualization of the electron velocity distribution function (VDF) in the selected box, shown in $(v_\parallel, v_{\perp,1}, v_{\perp,2})$ space. The VDF exhibits a clear elongation along the magnetic-field direction and a double-core structure. 
    Top row: Two-dimensional projections of the VDF in the $(v_\parallel, v_{\perp,1})$ and $(v_\parallel, v_{\perp,2})$ planes. 
    Middle row: Spatial maps of electron agyrotropy ($\sqrt{Q_e}$), parallel electron heating ($T_{\parallel,e}/T_{\parallel,e0}$).
    Bottom row: Electron mixing fraction ($F_e$), and $\log_{10}|J_{z,e}|$, with the analyzed subregion highlighted by the magenta box.}
    \label{fig:appendix_vdf_example}
\end{figure*}

To illustrate the phase-space structure associated with localized electron energization, we examine a representative subregion within the lower shear layer during the nonlinear stage ($t=723\Omega_{c,i}^{-1}$). The region is selected based on the spatial diagnostics shown in \autoref{fig:heating_maps_spectra} and exhibits a localized increase in $T_{\parallel,e}/T_{\parallel,e0}$, accompanied by moderate agyrotropy ($\sqrt{Q_e}$) and enhanced out-of-plane current density ($|J_{z,e}|$) (see \autoref{fig:appendix_vdf_example}). The mixing fraction $F_e$ (defined in \citealt{nakamura2014}), which quantifies the relative contribution of distinct plasma populations, indicates that the region is dominated by a single plasma population, excluding the possibility that the observed features arise from the superposition of distinct inflowing populations.

The velocity-space distribution function measured within the highlighted box shows clear deviations from a Maxwellian shape, including elongation along the local magnetic-field direction and a distinct double-core structure. These features are consistent with strong parallel energization and the development of non-Maxwellian phase-space structures in regions of intense current activity. The absence of strong mixing signatures indicates that they arise from localized field--particle interactions and pressure-tensor effects rather than from population mixing.

\end{document}